\documentclass{PoS}
\usepackage{epsfig}
\def \beq{\begin{equation}}
\def \eeq{\end{equation}}
\def \beqa{\begin{eqnarray}}
\def \eeqa{\end{eqnarray}}
\def \Ds{\ensuremath{D_s}}
\def \Dt{\ensuremath{D_{\tau}}}

\def \Z{{\cal Z}}
\def \C{{\cal C}}
\def \cs{\ensuremath{c_s}}
\def \cv{\ensuremath{c_{\scriptscriptstyle V}}}
\def \lambdams{\Lambda_{\overline{\scriptscriptstyle MS}}}
\def \etal{{\sl et al.\/}}

\def \np{{\sl Nucl.\ Phys.\/}}

\def \pr{{\sl Phys.\ Rev.\/}}
\def \prl{{\sl Phys.\ Rev.\ Lett.\/}}
\PoS{PoS(LAT2005)173}

\title{A new method for computation of QCD thermodynamics: EOS, specific
       heat  and speed of sound
       \\ \hfill {\small \sl TIFR/TH/05-38}\\ \vspace{-0.5cm}}

\ShortTitle{A new method for computation of QCD thermodynamics} 

\author{Rajiv V. Gavai\\ 
        Department of Theoretical Physics,\\
        Tata Institute of Fundamental Research\\
        Homi Bhabha Road, Mumbai 400005, India. \\
        E-mail:~\email{gavai@tifr.res.in}}

\author{Sourendu Gupta\\
        Department of Theoretical Physics,\\
        Tata Institute of Fundamental Research\\
        Homi Bhabha Road, Mumbai 400005, India. \\
        E-mail:~\email{sgupta@tifr.res.in}}

\author{\speaker{Swagato Mukherjee}
        \thanks{The speaker would like to thank Lattice 2000 organizing 
                committee and TIFR for providing the necessary financial 
                support to attend this conference} \\
        Department of Theoretical Physics,\\
        Tata Institute of Fundamental Research\\
        Homi Bhabha Road, Mumbai 400005, India. \\
        E-mail:~\email{swagato@tifr.res.in}}

\abstract{We propose a new variant of the operator method for the
computation of the equation of state of QCD, which yields positive
pressure for all temperatures and all values of temporal lattice
spacings. Using this new method, we calculate the continuum limit of
pressure, $P$, energy density, $\epsilon$, entropy density, $s$,
specific heat, \cv, and the speed of sound, \cs, in quenched QCD, for
$0.9 \le T/T_c \le 3$.}

\FullConference{XXIIIrd International Symposium on Lattice Field Theory\\

                 25-30 July 2005\\

                 Trinity College, Dublin, Ireland}

\begin{document}
\section{Introduction} \label{sc.intro}
The equation of state (EOS) of quantum chromodynamics (QCD) has been
perused for a long time, not only because it is theoretically
interesting but also for its growing importance in the relativistic
heavy-ion collision experiments. Almost two decades back, an operator
method formalism was devised \cite{engels} to study the EOS of QCD
numerically by using lattice gauge theory. Later it was found that
computations on coarse lattices, using this method, gave negative
pressure near the phase transition region. It was argued then, that this
problem arises due the use of perturbative formulae for the various
derivatives of the coupling. To cure this problem a new elegant method,
the integral method, was proposed in \cite{boyd}. This method assumes
the system to be homogeneous, bypassing the use of those derivatives. It
uses another non-perturbatively determined derivative of the coupling,
namely the QCD $\beta$-function.  Unfortunately, the assumption of
homogeneity does not hold at a first order phase transition.  Moreover,
the method also has other limitations like the pressure being
ill-defined below an arbitrarily chosen cut-off temperature and the
fluctuation measures being obtained only by taking numerical
derivatives, resulting in large errors.

In view of this situation we propose \cite{swagato2} a new variant of
the operator method which gives non-negative, well-defined pressure for
the entire temperature range. This method can also be extended
to the calculation of fluctuation measures \cite{swagato1}, starting
from the first principle. We choose the temporal lattice spacing to set
the scale of the theory, in contrast to the choice of the spatial
lattice spacing in the approach of \cite{engels}. Thus, our
method \cite{swagato2} could be called the t-favoured operator method
and the method of \cite{engels} may be called the s-favoured operator
method. 
\section{Formalism} \label{sc.formalism}
As in most finite temperature lattice gauge theory formulation, our
t-favoured formalism is defined on a $3+1$ dimensional hypercubic,
asymmetric lattice having different lattice spacings $a_{\tau}$ and
$a_s$ in the temporal and spatial directions respectively. The
temperature($T$) and the spatial volume($V$) are defined as
$T=(a_{\tau}N_{\tau})^{-1}$ and $V=(a_sN_s)^3$, $N_{\tau}$ and $N_s$
being the number of lattice sites in the temporal and spatial
directions. In our t-favoured scheme we introduce the anisotropy
parameter $\xi$ and the scale $a$ by the relations---
\beq
  \xi=\frac{a_s}{a_{\tau}},
  \qquad{\rm and}\qquad
  a=a_{\tau}.
\eeq
The partial derivatives with respect to temperature and volume can
easily be written in terms of these variables---
\beq
  T\left.\frac{\partial}{\partial T}\right|_{V} =
    \xi\left.\frac{\partial}{\partial \xi}\right|_{a} -
      a\left.\frac{\partial}{\partial a}\right|_{\xi},
  \qquad{\rm and}\qquad
  V\left.\frac{\partial}{\partial V}\right|_{T} =
    \frac\xi 3\left.\frac{\partial}{\partial \xi}\right|_{a}.
\label{eq.lat-der}    
\eeq

In finite temperature lattice gauge theory simulations the scale is set
by the temperature $T=1/a_{\tau}N_{\tau}$. In that sense our choice of
scale $a=a_{\tau}$ corresponds to the situations in actual simulations, giving
rise to the most natural procedure for the scale setting. On the other
hand, in the case of the s-favoured scheme \cite{engels}, for any
$\xi\ne1$, the scale is set by $a=a_s$ and only in the $\xi\to1$ limit
this natural choice of the scale emerges.

For a pure gauge $SU(N_c)$ theory on an asymmetric lattice, the Wilson
action and the corresponding partition function are defined as---
\beq
  S[U]= K_{s}P_s+K_{\tau}P_{\tau},
  \qquad{\rm and}\qquad
  \Z(V,T)=\int{\cal{D}}U e^{-S[U]},
\label{eq.action}  
\eeq
where periodic boundary conditions are assumed. $P_s$ denotes the sum of
spatial plaquettes over all lattice sites, and $P_{\tau}$ is the
corresponding sum of mixed space-time plaquettes. The couplings in the
temporal and spatial directions are given by
$K_s=2N_c/\xi g_s^2$, and $K_{\tau}=2N_c\xi/g_{\tau}^2$ respectively.

In the weak coupling limit, $g_i^{-2}$ 's ($i=s,\tau$) can be expanded
\cite{hasenfratz} around their symmetric lattice value $g^{-2}(a)$,
\beq 
   g_i^{-2}(a,\xi)=g^{-2}(a)+c_i(\xi)+O[g^2(a)],
\label{eq.coupling}
\eeq
with the condition $c_i(\xi=1)=0$. The value of $g^2(a)$ depends on the
identification of the scale $a$. 

We are now in a position to calculate different thermodynamic
quantities. Let us first look at the energy density ($\epsilon$) and the
pressure ($P$). Starting from the standard definitions of $\epsilon$ and
$P$ in terms of $\ln\Z$, operating the lattice derivatives of
eq.~(\ref{eq.lat-der}) on the partition function given in
eq.~(\ref{eq.action}) and making use of the relation mentioned in
eq.~(\ref{eq.coupling}) one can  derive the expressions for $\epsilon$
and $P$. In the $\xi\to1$ limit these expressions reduces to---
\beqa  
  \frac\epsilon{T^4} &=& 6N_cN_\tau^4 \left[\frac{\Ds-\Dt}{g^2}
    -(c_s'\Ds+c_{\tau}'\Dt) \right] + 
    6N_cN_\tau^4 \frac{B(\alpha_s)}{2 \pi \alpha_s^2} \biggl
    [\Ds+\Dt\biggr]
  \qquad{\rm and}\qquad \nonumber \\ 
  \frac{P}{T^4} &=& 2N_cN_\tau^4 \left[\frac{\Ds-\Dt}{g^2}
    -(c_s'\Ds+c_{\tau}'\Dt) \right].
\label{eq.e-p1}
\eeqa
Here the primes denote derivatives with respect to $\xi$, $B(\alpha_s)$
is the QCD $\beta$-function, with the usual definition of
$\alpha_s=g^2(a)/4\pi$. An ultraviolet divergence in the values of the
plaquettes is removed by a subtraction of the corresponding vacuum
($T=0$) values to yield $D_i=\langle P_i\rangle -\langle P_0\rangle$
above. In \cite{karsch},  $c_i'$ 's have been calculated for a weakly
coupled $SU(N_c)$ gauge theory upto one-loop order. Since these Karsch
coefficients are known only upto one-loop order, we also use one-loop
order $B(\alpha_s)$ for logical consistency. We use the one-loop order
renormalized coupling $g^2(a)$ as has been suggested in \cite{sourendu}. 

Our expressions for energy density and pressure, eq.~(\ref{eq.e-p1}),
can be compared to the expression of $\epsilon$ and $P$ in the
s-favoured operator method of \cite{engels}. It is clear that the $P$ in
the t-favoured method is exactly $\epsilon/3$ of the s-favoured method.
The positivity of the $\epsilon$ in the s-favoured scheme guarantees the
positivity of $P$ in the t-favoured scheme. The interaction measure--- 
\beq
  \frac{\Delta}{T^4}=\frac{(\epsilon-3P)}{T^4}=6N_cN_\tau^4 
    \frac{B(\alpha_s)}{2 \pi \alpha_s^2}\biggl[\Ds+\Dt\biggr],
\label{eq.delta}      
\eeq
is same for both the t and s-favoured methods. Since $\Delta$ is always
positive our expression for $\epsilon$ is also bound to give positive
values for the energy density.

Let us now turn our attention to specific heat at constant volume
($\cv$) and the isentropic speed of sound ($\cs$). These quantities are
defined as--- 
\beq 
  \cv = \left.\frac{\partial\epsilon}{\partial T}\right|_V, 
  \qquad{\rm and}\qquad 
  \cs^2 \equiv \left.\frac{\partial P}{\partial\epsilon}\right|_s = 
    \left.\frac{\partial P}{\partial T}\right|_V \left(\left.
    \frac{\partial\epsilon}{\partial T}\right|_V\right)^{-1} = 
    \frac{s/T^3}{\cv/T^3}.  
\label{eq.cv-cs} 
\eeq
Here we have used the definition of entropy density, $s=(\partial S
/\partial V)_T=(\epsilon+P)/T$, and the thermodynamic identity
$(\partial P/\partial T)_V=(\partial S/\partial V)_T$, $S$ being the
total entropy. In \cite{swagato1} it has been argued that the specific
heat can most easily be obtained by working in terms of the
dimensionless variable $\C$, the so-called conformal measure, --- 
\beq
  \C = \frac\Delta\epsilon, 
  \qquad{\rm and}\qquad \Gamma = T\left.\frac{\partial\C}{\partial T}
    \right|_V.  
\label{eq.gamma} 
\eeq 
By doing some straight forward algebraic manipulations it is very easy
to see that--- 
\beq 
  \frac\cv{T^3} = \left(\frac{\epsilon/T^4}{P/T^4}\right)
    \left[\frac s{T^3}+\frac{\Gamma}{3} \frac\epsilon{T^4}\right],
  \qquad{\rm and}\qquad 
  \cs^2= \left(\frac{P/T^4}{\epsilon/T^4}\right)
    \left[1+\frac{\Gamma\epsilon/T^4}{3s/T^3}\right]^{-1}.
\label{eq.cv-cs1}
\eeq 
In \cite{swagato2}, an  expression for $\Gamma$ has been calculated in
terms of the plaquettes and the co-variances of the plaquettes.  It has
also been shown in \cite{swagato1} that $\cv/T^3$ and $\cs^2$ reaches
their respective ideal gas values in the $g\to0$ limit. The expression
for $\Gamma$ contains the second derivatives of the Karsch coefficients
$c_i''$ 's.  Numerical values of these second derivatives have already
been evaluated in \cite{swagato1} for an $SU(3)$ gauge theory. Using all
these we can determine the specific heat and the speed of sound directly
from numerical simulations. 
\section{Results}\label{sc.results}

We have performed simulations in the temperature range $0.9 \le T/T_c
\le 3$, with temporal lattice sizes $N_{\tau}=8,~ 10,~12$. Spatial
lattice sizes were chosen to be $2N_{\tau}+2$ for $T\le2T_c$, and
$3N_{\tau}+2$ for $3T_c$. For the zero temperature simulations, a
minimum lattice size of $22^4$ were chosen and then lattice sizes were
scaled up with changes in lattice spacings. Typical number of sweeps
used for the measurements of the plaquettes were around few hundred
thousand in finite temperature runs, and around one hundred thousand in
zero temperature simulations. The details of our simulations can be
found in \cite{swagato2}. 

\begin{figure}[!ht]
\begin{center}
  \includegraphics[scale=0.55]{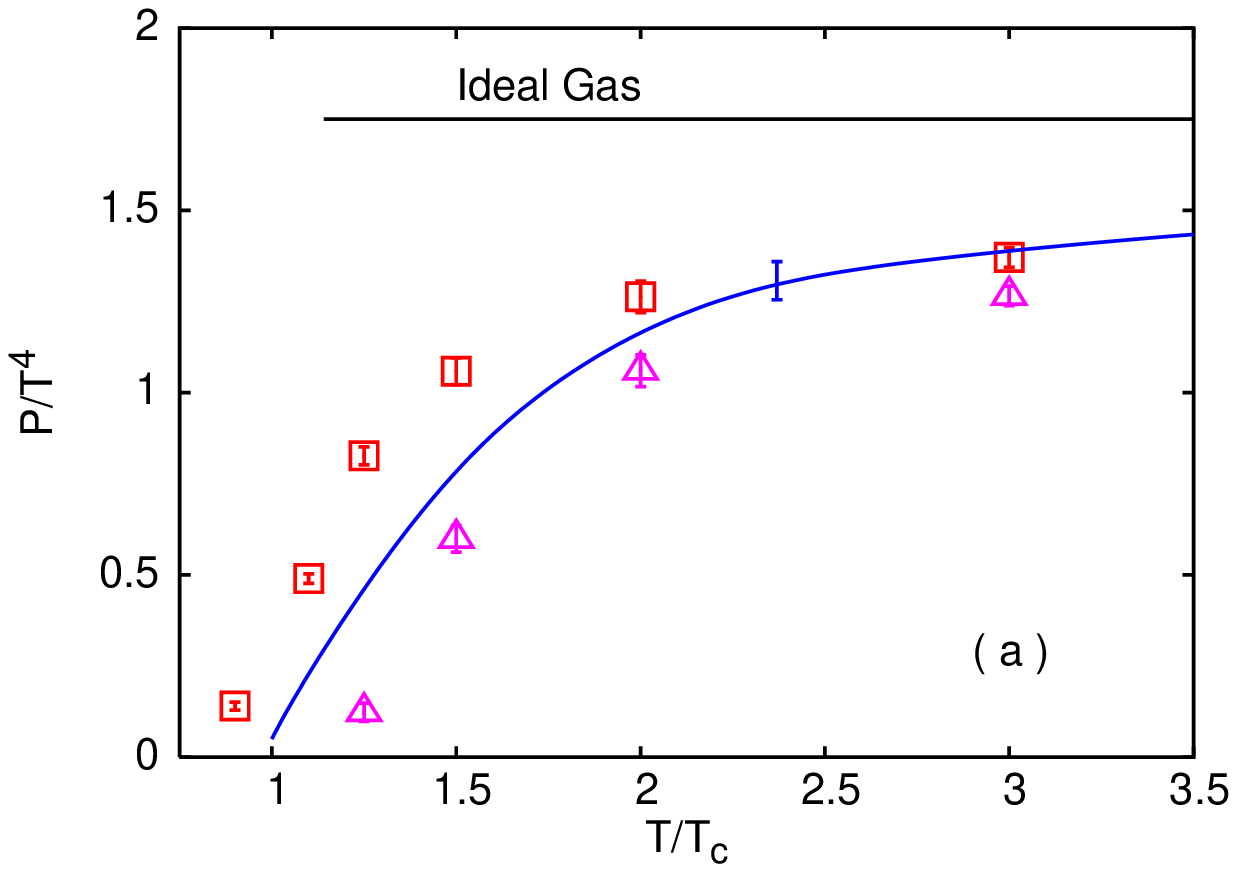}
  \includegraphics[scale=0.55]{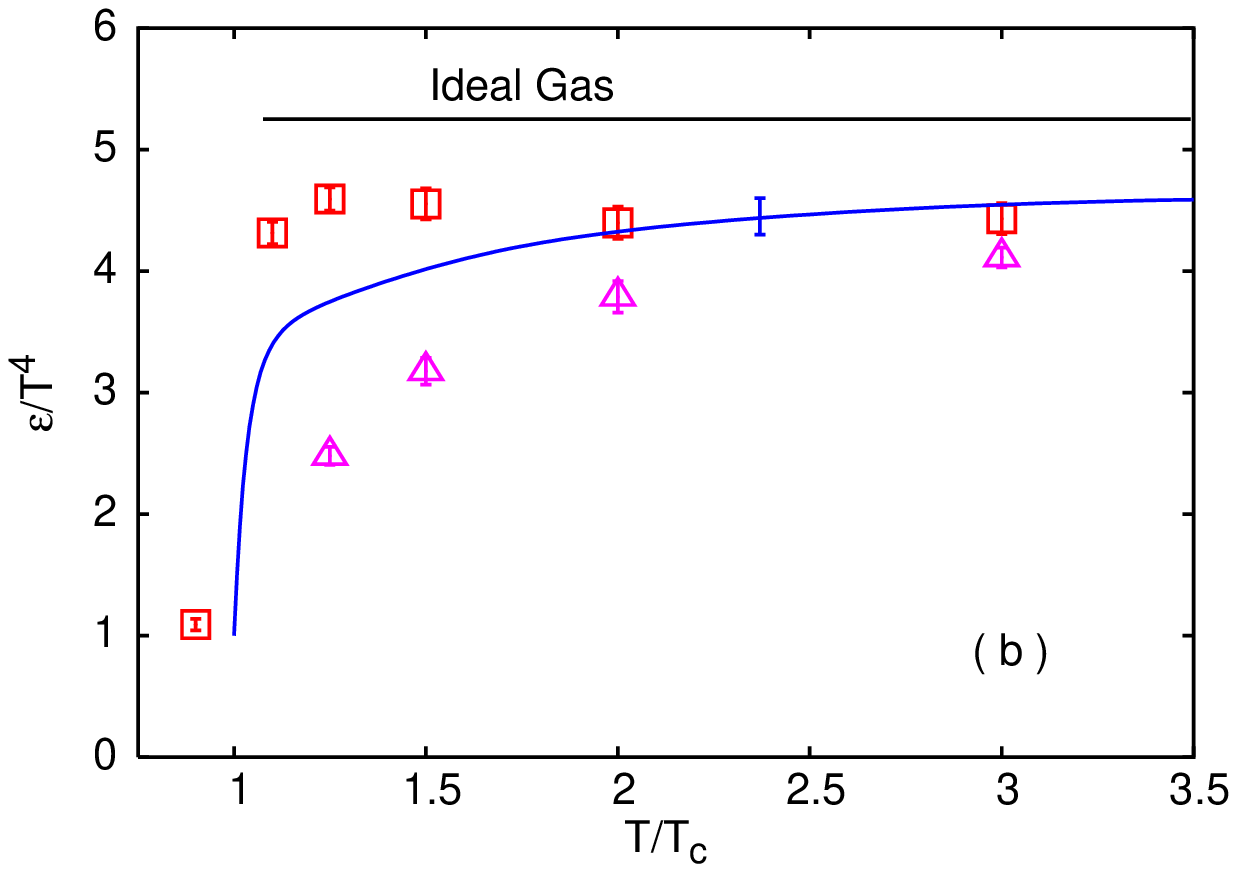}
  \includegraphics[scale=0.55]{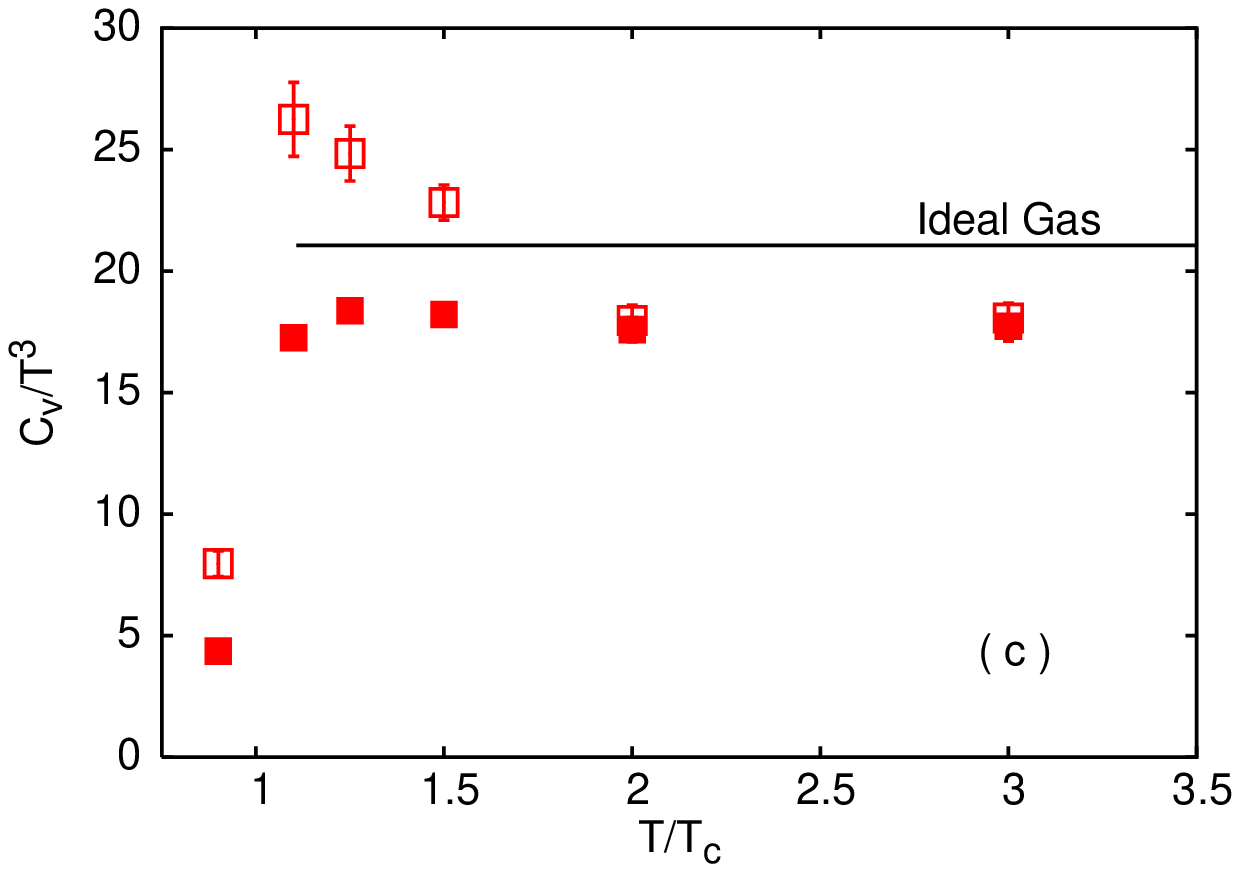}
  \includegraphics[scale=0.55]{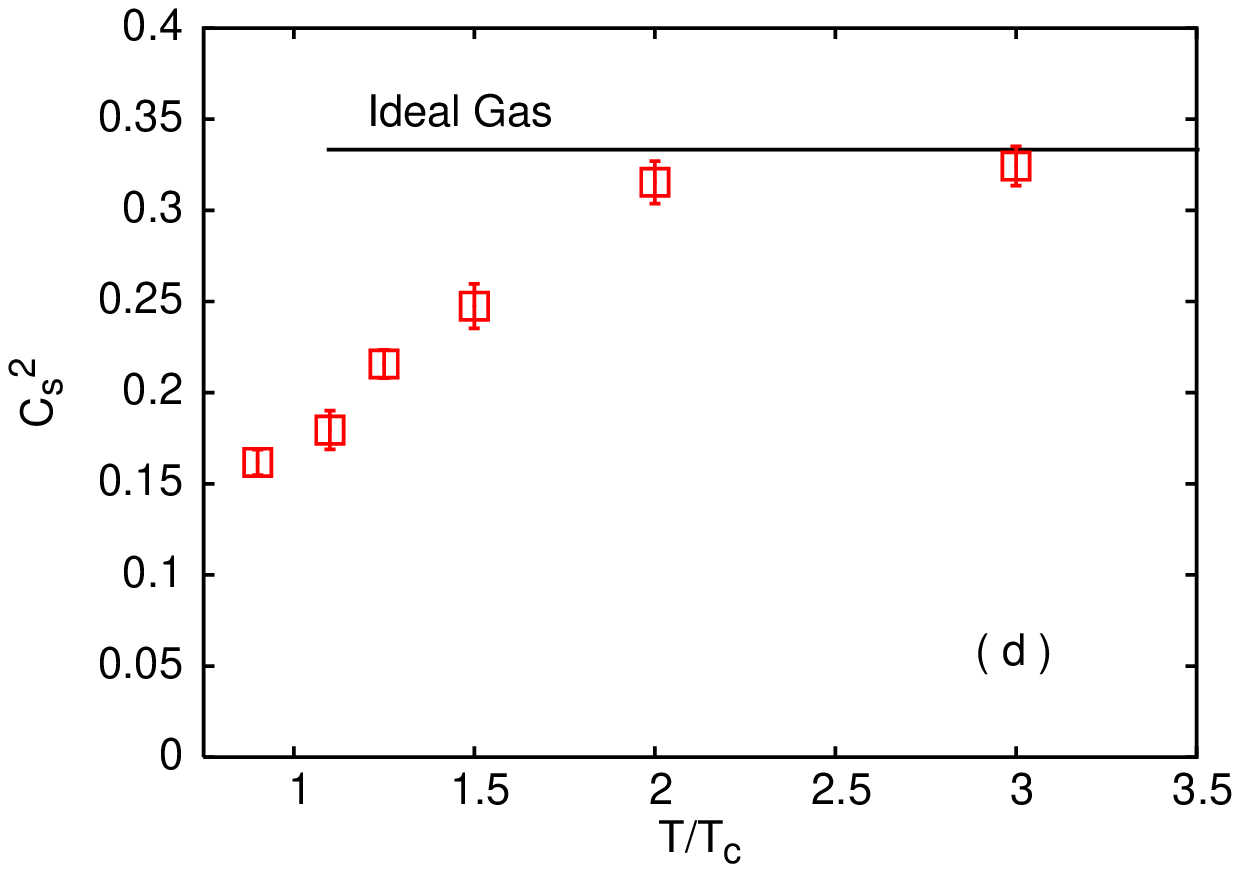}
  \caption{We show comparisons between the continuum results of
  different thermodynamic quantities for our new method (boxes), the old
  operator method (triangles) and the integral method (line). In panel
  (c) we show a comparison between our continuum results for $\cv/T^3$
  (open boxes) and continuum $4\epsilon/T^4$ (filled boxes). The data
  for the integral method has been taken form Ref. \cite{boyd}.}
\label{fig.cont}
\end{center}
\end{figure}

From the computations for the above mentioned three temporal lattices
we performed continuum extrapolations by linear fits in $a^2
\propto 1/N_{\tau}^2$ at all temperatures. In Figure~(\ref{fig.cont}a)
and Figure~(\ref{fig.cont}b) we compare the continuum limits of pressure
and energy density for the t-favoured method \cite{swagato2}, s-favoured
method \cite{engels} and the integral method \cite{boyd}. The most
important thing to be noted is that, unlike the old s-favoured operator
method, the new t-favoured operator method gives positive pressure for
all the temperatures, and also for all lattice spacings. The t-favoured
method pressure agrees well with the integral method pressure in the
high temperature ($T\ge2T_c$) region and both differ from the ideal gas
value by about $20\%$. In contrast, even in the continuum limit, and on 
$N_{\tau}$ as large as $8-12$, the s-favoured method does not yield  
positive pressure near $T_c$. On the other hand in the phase transition 
region the t-favoured method pressure shows a steeper rise than that of the
integral method pressure. Compared to the integral method the energy
density in the t-favoured method is harder near $T_c$ and agrees with
that of the integral method for $T\ge2T_c$.  This indicates a difference
in the latent heat determined by the two methods. 

Part of the disagreement between the t-favoured and the integral method
results can be traced back to the fact the integral method pressure is
simply set to zero at $T_c$ and below this cut-off temperature the integral
method pressure is ill-defined. Also the assumption of a homogeneous
system breaks down at a first order phase transition that takes place
for a pure $SU(3)$ gauge theory. Along with these, the integral method
results of \cite{boyd} shown here are for coarser lattices, uses
non-perturbative $\beta$-function and different scheme to determine the
renormalized coupling. We plan to investigate these issues in future, in
view of all the differences in these two computations.

We present the continuum results for the specific heat and the speed of
sound in Figure~(\ref{fig.cont}c) and Figure~(\ref{fig.cont}d)
respectively. For $T\ge2T_c$, $\cv/T^3$ is far from its ideal gas value,
but is quite consistent with the prediction in conformal theories that
$\cv/T^3=4\epsilon/T^4$. In the neighbourhood of $T_c$, $\cv$ shows a
peak-like structure. Whereas $\cs$ is consistent with its ideal gas
value (in accordance with the prediction for a conformal theory) for
$T\ge2T_c$, it decreases dramatically near $T_c$.

\begin{figure}[!ht]
\begin{center}
  \includegraphics[scale=0.55]{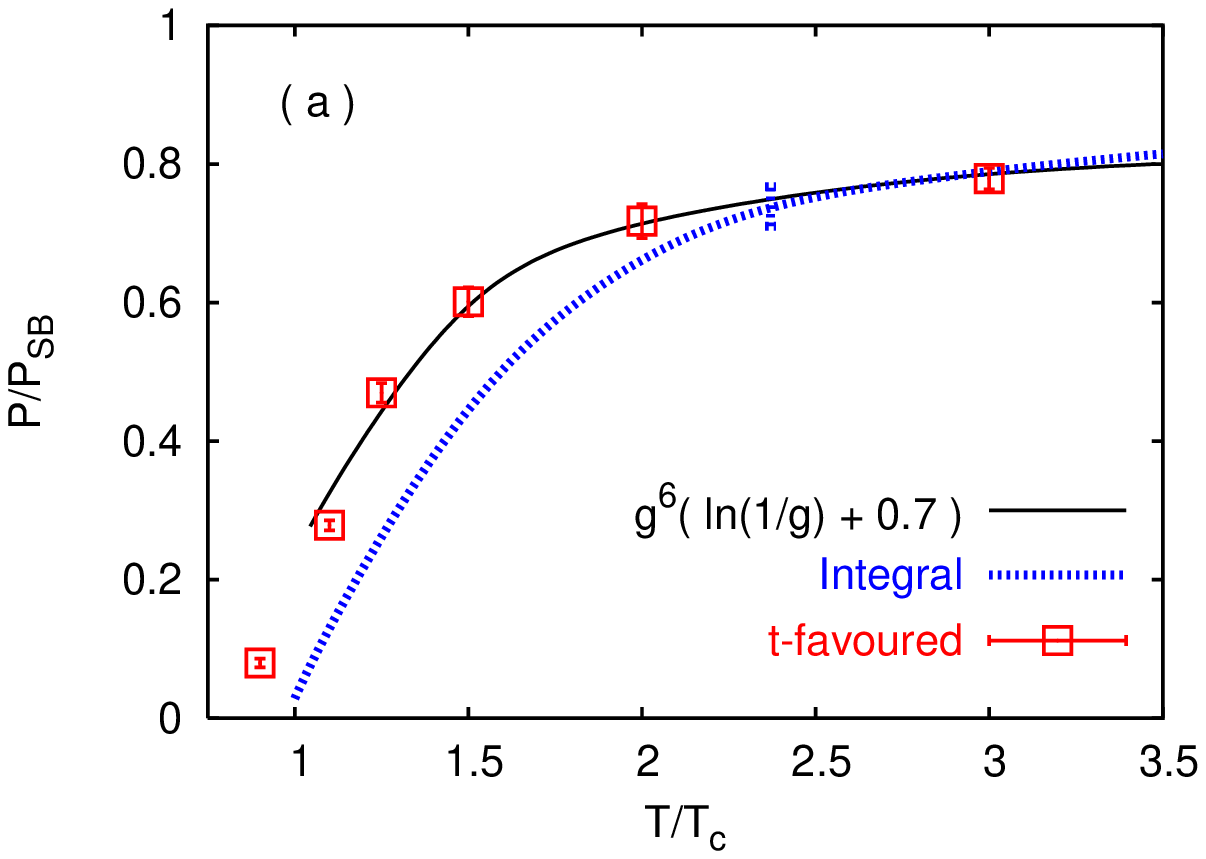}
  \includegraphics[scale=0.55]{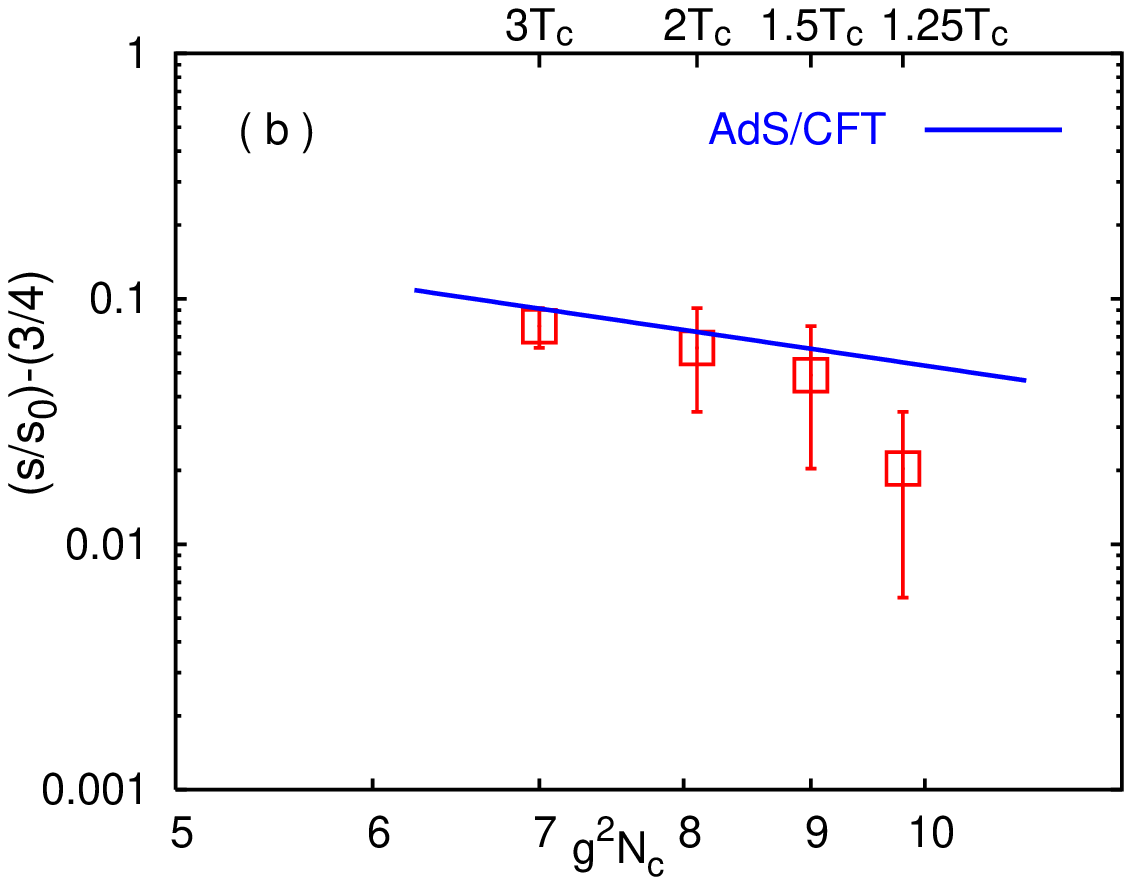}
  \caption{In panel (a) we compare the pressure obtained by t-favoured
  method (boxes), integral method (dotted line) and the $g^6\ln(1/g)$
  order perturbative expansion (solid line).  The data for the integral
  method and the perturbative expansion are taken from Ref.  \cite{boyd}
  and Ref.  \cite{kajantie} respectively. The values of the
  $T/\lambdams$ in Ref. \cite{kajantie} has been converted to $T/T_c$
  using the $T_c/\lambdams$ quoted in \cite{sourendu}. In panel (b) we
  show the deviation of $s/s_0$ from $3/4$ (boxes) as a function of the
  t'Hooft coupling. The line shows the same quantity as predicted by the
  AdS/CFT correspondence
\cite{klebanov}.}
\label{fig.pert-cft}
\end{center}
\end{figure}

We find that the pressure in the t-favoured method agrees with that
obtained from a dimensionally reduced theory, matched with the 4-d
theory perturbatively upto order $g^6\ln(1/g)$ \cite{kajantie}, almost
all the way down to $T_c$. This has been shown in
Figure~(\ref{fig.pert-cft}a).

In Figure~(\ref{fig.pert-cft}b) we compare our result for the entropy
density $s$ with that of the strongly coupled ${\cal N}=4$ Super
Yang-Mills (SYM) theory obtained by using the AdS/CFT correspondence
\cite{klebanov}. We find that the deviation of $s/s_0$, $s_0$ being the
entropy density for an ideal gluon gas, from $3/4$ is in good agreement
with the predicted results for a strongly coupled ${\cal N}=4$ SYM from
$T>1.5T_c$. This, along with our results for $\cv/T^3$ and $\cs^2$,
gives a hint that QCD behaves like a quasi-conformal theory for t' Hooft
couplings $g^2N_c<9$.
\section{Summary} \label{sc.summary}
We presented a new variant of the operator method, namely the
t-favoured scheme, for the calculation of the EOS of QCD. This methods
gives positive pressure for all temperatures and all lattice spacings,
even when the old operator method gives negative pressure. Using this
new method we have computed the continuum limit of the energy density
and pressure. We have also extended this method to determine the
continuum limit of specific heat at constant volume and speed of sound,
both above and below the transition temperature. To the best of our
knowledge this is the first continuum results for these two quantities.
We have also compared our results with the predictions of perturbation
theory and also with the results for strongly coupled ${\cal N}=4$ SYM.
We find good agreement with both in certain ranges of temperature.
\end{document}